 \documentstyle[aps,prd]{revtex}
\begin{document}
\draft

\title{The Head-On Collision of Two Equal Mass Black Holes}

\author{Peter Anninos${}^{(1)}$, David
Hobill${}^{(1,3)}$, Edward Seidel${}^{(1,2)}$, Larry Smarr${}^{(1,2)}$ and
Wai-Mo Suen${}^{(4)}$}

\address{${}^{(1)}$ National Center for Supercomputing Applications \\
605 E. Springfield Ave., Champaign, IL 61820}

\address{${}^{(2)}$ Department of Physics \\
University of Illinois, Urbana, IL 61801}

\address{${}^{(3)}$ Department of Physics and Astronomy \\
University of Calgary, Calgary, Alberta, Canada T2N 1N4}

\address{${}^{(4)}$ Department of Physics \\
Washington University, St. Louis, Missouri, 63130}

\date{\today}
\maketitle
\begin{abstract}
We study the head-on collision of two equal mass, nonrotating black
holes. Various initial configurations are investigated, including
holes which are initially surrounded by a common apparent horizon to
holes that are separated by about $20M$, where $M$ is the mass of a
single black hole. We have extracted both $\ell = 2$ and $\ell=4$
gravitational waveforms resulting from the collision.  The normal
modes of the final black hole dominate the spectrum in all cases
studied.  The total energy radiated is computed using several
independent methods, and is typically less than $0.002 M$.
We also discuss an analytic approach to estimate the total
gravitational radiation emitted in the collision by generalizing
point particle dynamics to account for the finite size
and internal dynamics of the two black holes.
The effects of the tidal
deformations of the horizons are analysed using the membrane paradigm
of black holes. We find excellent agreement between the numerical
results and the analytic estimates.
\end{abstract}



\section{Introduction}
\label{sec:introduction}

The spiralling coalescence of two black holes in orbit
about one another is considered to be one of the most promising
sources of gravitational waves~\cite{Thorne94a}. The strong burst of
gravitational waves resulting from such an event should be detectable
by the next generation of gravitational wave detectors such as LIGO
and VIRGO~\cite{LIGO3}. From observing these violent events in our
universe, we expect to obtain important insights into astrophysics,
gravitation and cosmology.  In particular, such signals should provide
the first direct and unambiguous evidence for the existence of black
holes if the unique signature of the quasinormal
modes~\cite{Chandrasekhar83} are excited. The information gained from
the detected waveforms should allow one to reconstruct the
astrophysical parameters of the system, such as the masses, spin, and
orbital angular momentum and linear momentum of the colliding black
holes, and the final black hole. Since LIGO and VIRGO are expected to
begin taking data by the end of the decade, it is important to perform
accurate calculations of the waveforms emitted during these
events. Numerically generated waveform templates will be essential for
the analysis of data collected by gravitational wave detectors.

In a series of papers~\cite{Anninos93b,Anninos93a,Anninos94a}
we investigate a special case of the black hole
coalescence problem, namely the head-on collision of two black
holes. On the one hand, the simplifying assumption of a head-on
collision reduces the general three dimensional coalescence problem to
a two dimensional axisymmetric problem, and is hence much more
tractable. On the other hand, a head on collision can be regarded as
an approximation to the last nonlinear stage of inspiralling
coalescence--the final plunge.

Our work extends and refines the earlier calculations of Dewitt, Cadez,
Smarr, and Eppley~\cite{Cadez71,Smarr75,Eppley75,Smarr76,Smarr77,Smarr79}
(henceforth abbreviated as DCSE).
Results from that collective body
of work suggest that the normal modes of the final black hole resulting
from the collision are excited and that the total energy released is
typically less than 0.1\% of the mass of the final black hole.
However these numerical calculations proved to be very difficult due
to inherent coordinate singularities and
numerical instabilities that plague the two black hole
system.
Also, the computer power available at that time did not permit
highly resolved evolutions, and they did not have waveform
extraction techniques~\cite{Abrahams90} at their disposal for
determining gauge-invariant waveforms.
For these reasons,
DCSE quote their results as uncertain
to within a factor of two, for example, in the total radiated energy
{}~\cite{Smarr79}.  It is therefore imperative to revisit this
important physical problem with the benefit of more powerful computers
and improved analytic and numerical techniques developed over the
intervening 15 years to calculate unambiguous waveforms and energy
fluxes resulting from the collision.

Building on the work of DCSE,
and more recent work involving distorted single black
holes~\cite{Anninos93c,Bernstein93b,Abrahams92a}, many of the
numerical problems associated with colliding two black holes
have been overcome in the present work.
In particular,
we have used a hybrid set of coordinates to
resolve the axis and saddle point problems encountered by DCSE
so that our evolutions are more accurate and more stable.
The numerical code we have developed can
evolve black holes with initial separation distances
between $\sim 4M$ and $\sim 20M$,
where $M$ is the mass parameter defined to
be half the ADM mass of the system.  (We note that $M$ is simply a
convenient parameter to characterize the coordinates and only
approximates the mass of a single black hole in the limit that the
holes are widely separated.)
We have applied more modern analysis techniques on the numerical
data to extract
waveforms and compute the total energies emitted.
Using gauge-invariant waveform extraction we are now able to
determine highly accurate $\ell=2$ waveforms and,
for the first time $\ell=4$ waveforms.
We also present analytic estimates of the total energy radiated in the
collision by taking into account the finite size and internal dynamics
of the two black holes. Comparisons
of these results to the numerical solutions
are valuable not
just as a confirmation of our results, but more importantly,
they provide a physical understanding of the numerical data.

Sec.~\ref{sec:background} provides an overview of the
basic theoretical and computational methods used in this work.
A more complete discussion of the initial data, the choices of
coordinate systems, gauges, and the numerical methods we developed
to handle the problems associated with the evolution of two black
hole spacetimes can be found in a companion paper
{}~\cite{Anninos94a}.
In Sec.~\ref{sec:numeric} we present results from our numerical
studies.  First we investigate the highly nonlinear and dynamical near
field region by looking at the evolutions of the metric
data and apparent horizons.  We further discuss the timing
of the merging of the holes and its implications. Next we turn to the
far field and discuss the extraction of gravitational waveforms and
compute the total energy radiated from the collision process using
various radiation indicators.  In Sec.~\ref{sec:analytic}, we present
a semi-analytic approach to estimate the total energy radiated and
compare these results to the numerical calculations obtained in
Sec.~\ref{sec:numeric}.  We conclude with
Sec.~\ref{sec:conclusions} by summarizing our results and outlining a
program to extend this work to more general black hole interactions.

\section{Computational Framework}
\label{sec:background}

We use the 3+1 (ADM) formalism~\cite{Arnowitt62} to write the
Einstein equations as a first order (in time) set of differential
equations
for the dynamical variables $\gamma_{ij}$ and
$K_{ij}$, the spatial 3-metric and extrinsic curvature respectively.
The maximal slicing condition $Tr~K=0$ is imposed throughout the
evolution. Also, because the spacetimes we work with possess
an axial Killing vector ($\partial/\partial x^3 \equiv \partial
/\partial\phi$) all variables are independent of the
azimuthal angle $\phi$.

This work, like that of DCSE, is based upon studying the
axisymmetric evolution of the analytic
Misner initial data~\cite{Misner60} representing two
equal mass black holes at the moment of time symmetry ($K_{ij}=0$).
The spatial 3-metric for this
data set can be written using cylindrical coordinates as
\begin{equation}
dl^2=\Psi_M^4\left(d\rho^2+dz^2+\rho^2d\phi^2\right),
\label{mmetriczr}
\end{equation}
where
\begin{equation}
\Psi_M=1+\sum_{n=1}^{\infty}\frac{1}{\sinh(n\mu)}
       \left(\frac{1}{\sqrt{\rho^2+(z+z_n)^2}}
        +\frac{1}{\sqrt{\rho^2+(z-z_n)^2}}
       \right),
\label{mPsizr1}
\end{equation}
and $z_n=\coth (n\mu)$.
The free parameter $\mu$ determines the total ADM
mass of the spacetime and the proper distance between the two throats.
The effect of increasing $\mu$ is to set the two black holes
(centered along $\rho=0$, at $z=\pm \coth \mu$) further
away from one another and to decrease the total mass of the system.

The Misner data
consists of two throats connecting two isometric sheets.
The throats are spheres
on which boundary conditions relating the metric across the two
sheets may be imposed.  Since the natural boundaries (the throats and
a sphere surrounding the system far from the throats) do not lie along
constant ($z$,$\rho$) coordinates, it is useful to introduce the
body-fitted Cadez \cite{Cadez71} coordinates ($\eta$,$\xi$) where
$\eta$ is the logarithmic ``radial'' coordinate and $\xi$ the ``angular''
coordinate.
The advantage afforded by this set of coordinates
(shown in Fig.~\ref{fig:cadezgrid}) is that they are
spherical near the throats of the black holes and also far away in the
wave zone, thus allowing us to deal with throat boundaries and
asymptotic wave form extractions in a convenient way.
The disadvantage is that the coordinate
transformation introduces a singular saddle
point at the origin ($z=\rho=0$) that is not present in cylindrical
coordinates.  This creates certain numerical difficulties that require
special treatment as we discuss briefly in the remainder of this section.
The success of our methods depends critically on
utilizing both sets
of coordinate systems (cylindrical and Cadez) to advantage.

We have investigated a number of different numerical schemes to solve
the problem of colliding two black holes head-on.  The basic idea that
evolved from our investigations is to solve for the Cadez
metric components, which we write as
\begin{equation}
\gamma _{ij}= \Psi^4 \, \hat{\gamma}_{ij}= \Psi^4
\pmatrix{A & C & 0      \cr
C & B & 0       \cr
0 & 0 & D\sin ^{2}\xi \cr }
\label{3metric}
\end{equation}
in the coordinate order ($\eta$, $\xi$, $\phi$),
on the Cadez grid and use a shift vector to set
$C=\partial_t C=0$.
(We note that only the conformal metric components are evolved.
The conformal factor $\Psi^4 = \Psi^4_M/J$, where $J$ is the
Jacobian of the Cadez/cylindrical coordinate transformation,
remains constant in time.)
This choice for the shift vector has the advantage of a diagonal 3-metric
which helps to suppress the axis instability and
simplifies the equations of evolution and the extraction of invariant
gravitational waves in the far field.
Furthermore, with this approach it is possible to define
variables for the two black hole system that obey the same evolution
equations with similar boundary conditions as the single distorted
black hole code developed in previous
work~\cite{Anninos93c,Bernstein93b}.  In fact, the two black hole code
in its final incarnation evolved from the code we developed for
distorted axisymmetric single black hole spacetimes and much of the
discussion in~\cite{Anninos93c,Bernstein93b} is directly applicable here.

The difficulty with Cadez coordinates is the singular saddle
point located within the computational domain at the
origin $z=\rho=0$ (see Fig.~\ref{fig:cadezgrid}).
We evolve data near the saddle point by
taking advantage of the fact that the spacetime metric
components in cylindrical coordinates are smooth everywhere,
including the saddle point.
We can therefore define a
cylindrical coordinate ``patch'' to
evolve the cylindrical metric
and extrinsic curvature components
on the Cadez grid over regions near the saddle point.
The two sets of components,
Cadez and cylindrical, are evolved everywhere
independently of each other (except for the
coupling at the patch boundaries) on a single Cadez grid. The
nonsingular cylindrical components are then used to correct the singular
Cadez components in the patched region using the
general tensor relations $T'_{ij}=(\partial
x^k/\partial x'^i)(\partial x^l/\partial x'^j) T_{kl}$.
The Cadez components, in turn,
provide corrections to their cylindrical counterparts
everywhere else, helping to suppress the
axis instability that is inherently present in the cylindrical
coordinate system possessing a nondiagonal metric.
A more detailed discussion of this procedure can be found
in~\cite{Anninos94a}.

Our code was subjected to a number of tests, including
matching waveforms to perturbation theory as
we do in Sec.~\ref{subsubsec:gaugeinvariant}.
We also performed various convergence studies using
100 (27), 200 (35) and 300 (55) radial (angular) zones. We have
shown that the convergence rate for the total radiated energy
is quadratic in the grid spacing and, more specifically,
differences in the
dominant $\ell=2$ waveforms
between the 200 and 300 radial zone evolutions is on the
order of just a few percent.
We discuss the accuracy and reliability of our calculations
throughout this article when appropriate
and refer the reader to
Ref.~\cite{Anninos94a} for more details.

\section{Numerical Results}
\label{sec:numeric}

We have investigated six separate cases of the Misner two black hole
data sets corresponding to different values of $\mu$.  The physical
attributes of the initial data for these six cases are summarized in
Table~\ref{table:horizon} where we show $M=M_{ADM}/2$
({\it ie.,} half the ADM mass),
the proper distance between the two throats, and whether the
data contains a single global apparent or
event horizon surrounding
the two holes.
In this section, we categorically
discuss the dynamics of both the near and far field regions for the
various cases.

\subsection{Near Field}
\label{subsec:near}

\subsubsection{Spatial Metric}
\label{subsubsec:metric}

First we present results for the case $\mu = 1.2$, which is a data set
corresponding to two holes that have already merged initially. The
initial data contains an apparent horizon that encircles both
throats.  In Fig.~\ref{grr1.2} we show the conformal metric function
$\hat{\gamma}_{\eta \eta}=A$ at the coordinate time $t = 25M$.
Notice
that a sharp peak surrounding the hole is developing.  This peak
develops essentially spherically around the two throats from early in
the evolution, showing that the system behaves as a single black hole
from the outset. These results are similar to those observed in
studies of single throat spacetimes
{}~\cite{Anninos93c,Bernstein93b}. The reasons behind the
development of the peak is clear: As the coordinates are dragged into
the hole, the proper distance between radial grid points increases
rapidly towards the throat. However, as shown
in Fig.~\ref{lapse1.2b} as the lapse goes to zero in the
region near the throats, ``freezing'' all motions there. Hence we see
only the growth of the proper distance between grids in coordinate
time in the region near the horizon, developing a sharp peak.  Such
grid stretching effects present one of the main difficulties in
evolving black hole spacetimes in any numerical simulation utilizing a
singularity avoiding time slicing.  During the course of evolution,
the dramatic change in the radial metric function becomes increasingly
more difficult to resolve numerically, and the region of causal
dependence eventually becomes much smaller than the coordinate grid
spacing (as the radial metric function increases and the lapse
decreases there).

For comparison, we show in Fig.~\ref{lapse3.25}
the lapse for the case $\mu = 3.25$ where
the throats are much more separated, forming two black holes
initially.
The display corresponds to an early time ($t=22.5M$)
in the evolution of the
system where the two holes are acting essentially
independently of each other as they begin to
fall together, and the collapse of the lapse locally around the throat
reflects this fact.  After the holes begin to coalesce, the lapse
collapses spherically around both throats which
are contained within the final black hole.

\subsubsection{Horizons}
\label{subsubsec:horizons}

We have already observed that at late times the intrinsic geometry of
the apparent horizons oscillates at the normal mode frequency of the
final black hole. This aspect of apparent horizon dynamics is
discussed in detail in Ref.~\cite{Anninos93a} for three different
spacetimes including colliding black holes.  Here we look at horizons
in the context of distinguishing one from two black hole data sets and
estimating the mass energy of radiated gravitational waves.

One can define an effective mass of a black hole based on its apparent
horizon via the relation~\cite{Hawking73}
\begin{equation} M_{h} =
\sqrt{\frac{A_{h}}{16 \pi}}
\label{holemass}
\end{equation}
where $A_{h}$ is the intrinsic area of the apparent horizon. This
relationship gives a lower limit for the mass of the black hole, since
we know the apparent horizon should lie within or (in the case of
stationary black holes) coincide with the actual event horizon. When
the system begins to settle down Eq.~(\ref{holemass}) provides a good
estimate of the black hole mass as the apparent horizon will lie very
close to the event horizon.

In Fig.~\ref{fig:horizon} we show the evolution of the horizon mass
computed from Eq.~(\ref{holemass}) for two different cases.
The numerical data for the case $\mu=1.2$
is shown as a solid line.
The mass of the hole $M_{h}$ is normalized to units of the total ADM
mass of the spacetime, so ideally we should have $M_{h} <1$ for all
time.  However, because the horizon is always found near the peak of
$\hat\gamma_{\eta\eta}$, $M_h$ is extremely sensitive to the precise
position of the located horizon as discussed, for
example, in Ref.~\cite{Anninos93a}.
Small errors in the height and shape of the metric functions arising
from inadequate resolution can exaggerate errors in the area
of the horizon derived from these metric functions. A typical effect
of this problem is to overestimate the horizon mass $M_{h}$ after the
metric functions become too sharp for the grid to resolve. (This
effect has also been discussed in Ref.~\cite{Seidel92a} where an
apparent horizon boundary condition was used to circumvent this
problem.)

Also shown in Fig.~\ref{fig:horizon} is the result for
the case $\mu=2.2$. In this
case the initial horizons are the two distinct throats and contain
only about 79\% of the total mass of the spacetime.
Because we use a slicing condition in which the lapse is zero on
both throats, each throat remains a marginally trapped surface
throughout the evolution. We track this surface only until another
trapped surface forms across the equator
($z=0$) to surround both throats.
This new horizon, when it forms, contains essentially all the mass of
the spacetime except for a small amount carried by the
radiation. Ideally one should see the horizon mass leveling off just
below the total ADM mass of the system, with the difference being
accounted for by the energy carried away by gravitational
radiation. However the errors in the horizon mass, due to effects
discussed above, are large enough to hide this small amount of
radiation which is less than 0.1\% of the total mass.
Fig.~\ref{fig:horizon} also
shows a spurious feature appearing at the time of
the formation of the new horizon. This feature is due to numerical
difficulties with locking on to the new horizon which first appears
near the coordinate singularity present in the Cadez coordinate system.
After a brief period, the system settles to the final
black hole configuration (apart from the numerically induced slow
growth in the horizon mass as discussed above.)

The results for holes with wider initial
separations are similar to the case
$\mu = 2.2$, except that the time scale for the merging of the holes
is longer. Thus, by the time the holes have merged, as measured by the
appearance of an outer horizon, the horizon mass is overestimated by
larger amounts.  In the case $\mu=3.0$ this effect is about 20\%. Note
that the wavelength of the quasinormal mode of a black hole scales as
its mass, so this is consistent with the results of
Sec.~(\ref{subsubsec:gaugeinvariant}) which show that the extracted
waveform for these cases have wavelengths slightly longer than
expected for a final hole of mass $M_{ADM}$.

Although apparent horizons can confirm when two black holes have
definitely merged, the presence of two distinct apparent horizons does
not guarantee that the holes are separate, as a global event horizon
may still surround the individual holes.
The claim that the larger $\mu$ cases represent two distinct
black holes is supported by computing the area of a 2-sphere
defined on a constant $\eta$ surface that just encircles the
saddle point and therefore both holes.
In Fig.~\ref{fig:hoop}
we plot the area of this surface on the initial time slice
as a function of separation (and mass) parameter $\mu$.
The horizontal line is $16\pi$, the area corresponding to a
surface representing an effective mass equal to the ADM mass
(our units are such that the ADM mass is normalized to unity).
The point of intersection is $\mu \sim 2.1$,
and Fig.~\ref{fig:hoop} indicates that for larger
values of $\mu$ the data represents two black holes.

A more convincing and precise argument can be realized
by integrating light rays radially outward along the equator
starting from midway between the black holes.  If the photons ``escape'',
then clearly the system contains two disjoint event horizons initially
and hence two separate black holes.
(Whether photons escape or not is determined by their position
relative to the apparent horizon at late times,
typically $\sim 80M$. At such late times the apparent horizon
is expected to lie near the event horizon.)
We have performed several such studies in an attempt to find a
critical value for $\mu$ that separates the initial data into one or two
black hole sets.
Our results indicate that this value is approximately 1.8.
For $\mu \gtrsim 1.8$ the photons escape, while
for $\mu \lesssim 1.8$ they do not.
Furthermore, we have recently
developed a method for tracing out the actual
event horizon surface~\cite{Anninos94f} and found results
consistent with those obtained by integrating photons.

\subsubsection{Collision Timings}
\label{subsubsec:collision}

In Fig.~\ref{fig:time} we compare different timings
of the collision for each evolution
considered. The meaning of the various timings is as follows:
$t_{newt}$ is the Newtonian
free fall time for two point particles to collide from rest
and separated initially
by a distance $L/M$, $t_{hor}$ is the time at which the first merged
apparent horizon appears, and $t_{ring}$ is the time at which
the first (negative)
peak in the $\ell = 2$ gauge-invariant waveform reaches the
detector at $r = 40M$,
indicating the (retarded) onset of the quasinormal ringing
of the final black hole. First we note the remarkable coincidence
between the free-fall times to collision and the time required for the
apparent horizons to merge. This agreement between such different
indicators of the coalescence time is very satisfying.

Another interesting feature is the timing of the onset of the
quasinormal ringing. Note that for $\mu$ below 2.2
($L/M < 8.9$) the ringing begins
at about the same time regardless of the separation between the
holes.  For holes separated by more than this, a delay
in the onset of ringing becomes apparent. The fact that the ringing
begins at the same time regardless of their separation is not
surprising for $\mu$ below 1.8. As we discussed above, the two throats
are really just a single distorted black hole and the separation
between throats is physically irrelevant. For $\mu$ between 1.8 and
2.2, as the two holes are initially disjoint, one might expect a delay
in the onset of ringing. However, the important physical property of
the system governing the quasinormal ringing is the gravitational
scattering potential barrier surrounding the holes, not the position
of the event horizon itself. Since the peak of this potential barrier
is located near $r=3M$ for a Schwarzschild hole and the horizon is at
$r=2M$, we expect the potential barrier of each hole to merge into a
single one before the horizons do. Therefore, for black holes that are
initially close enough that their potential barriers have effectively
merged, we expect to see a system that behaves essentially like a
single distorted black hole in terms of quasinormal mode ringing, even
if the event horizons are distinct. From Fig.~\ref{fig:time}
it is clear that this transition from merged
potentials to distinct potentials takes place near $\mu \sim
2.2$.

\subsection{FAR FIELD}
\label{subsec:far}

\subsubsection{Gauge-Invariant Waveform Extraction}
\label{subsubsec:gaugeinvariant}

The main method we use to
calculate waveforms is based on the gauge invariant extraction
technique developed by Abrahams and Evans~\cite{Abrahams90} and
applied in Ref.~\cite{Abrahams92a} to black hole spacetimes.
The basic idea is to split the spacetime
metric into a spherically symmetric (static) background and a small
perturbation in the region where the curvature is dominated by the
mass content of a small compact object. We first expand the metric
perturbation in $m=0$ spherical harmonics $Y_{\ell 0}(\theta)$ and
their tensor generalizations. The Regge-Wheeler perturbation functions
are then extracted from the numerically computed metric components and
used to construct the gauge invariant Zerilli function $\psi$. (See
\cite{Abrahams90} for a detailed discussion of this procedure.) $\psi$
represents the wavelike part of the metric that is radiative at large
distances from the source and is commonly used in semi-analytic
calculations of black hole normal mode
frequencies~\cite{Chandrasekhar83}. The asymptotic energy flux carried
by gravitational waves can be computed from
\begin{equation}
\frac{dE}{dt} =\frac{1}{32\pi}
\left(\frac{\partial \psi}{\partial t}\right)^2, \label{zloss}
\end{equation}
independently for each $\ell$ mode contribution for the normalization
we use for $\psi$.

For all of the cases studied in this paper we have extracted both the
$\ell = 2$ and $\ell = 4$ waveforms at radii of $30, 40, 50, 60,$ and
$70M$.  (Coordinate positions corresponding to physical distances in
units of $M$ are approximated from the initial data in the
asymptotically spherical far field as $r\sim
\sqrt{\gamma_{\xi\xi}}/M=\Psi^2 /M$.)
By comparing results at each of these
radii we are able to check the propagation of waves and the
consistency of our energy calculations.

In Fig.~\ref{figl2mu1.2} we show the $\ell = 2$ waveform
(solid line) extracted at a
radius of $40M$ for the case $\mu=1.2$. This result is
similar to waveforms extracted from simulations of single distorted
black holes (see, for example, Ref.~\cite{Abrahams92a}) even though the
initial data sets and coordinate systems used are significantly
different. Of course since there is a horizon surrounding both
throats,
we expect this system to evolve as a single black hole from
the outside. As a perturbed single black hole system, we also expect
the quasinormal modes of the black hole to be excited. The dotted line
in Fig.~\ref{figl2mu1.2} shows the fit of
the lowest two (fundamental and first overtone) $\ell = 2$ modes of a
black hole of mass $(2M)$, over the range $70 < t/M < 160$,
obtained from
Refs.~\cite{Leaver75,Seidel90a}. The fit is excellent, showing that
the normal mode is the dominant part of the emitted radiation.  We
note that the first overtone quasi-normal mode is much more strongly
damped than the fundamental, and hence does not contribute appreciably
to the fit at late times.  Its main effect is to increase the accuracy
of the fit to the first peak in the extracted waveform.

Next we discuss the case $\mu = 2.2$, for which there are no initial
common apparent nor event horizons. In Fig.~\ref{figl2mu2.2} we show
the $\ell=2$ extracted waveform for this case. The solid line shows
the waveform detected at a distance $r=40M$ and the long dashed line
shows the waveform extracted at $r=60M$. The wave is clearly
propagating away from the hole at light speed with essentially
invariant shape and amplitude, with a wavelength of $2\times16.8M$,
confirming the original findings of Smarr and
Eppley~\cite{Smarr79}. However, our more accurate code now allows us
to go beyond estimating the wavelength and to fit quantitatively the
waveform to results known from black hole perturbation theory. The
short dashed line shows the result of fitting the $r=40M$ waveform (in
the range $64M<t<160M$) to a linear combination of the fundamental and
first overtone of the $\ell=2$ quasinormal mode for the final black
hole with mass $2M$. The fit is quite good, matching both the
wavelength and damping time, showing that the final black hole mass is
indeed very close to the total mass of the spacetime.

In Fig.~\ref{figl4mu2.2} we show the more difficult $\ell=4$ waveform
for the same case $\mu=2.2$, extracted at the same radius
$r=40M$. Again this waveform has been fit (over a similar range) to a
superposition of the fundamental and first overtone $\ell=4$
quasinormal modes of the black hole.  Although the fit to the
extracted waveform is reasonably good in terms of the wavelength and
damping time, this waveform is rather sensitive to the computational
parameters such as grid resolution and the extent of the numerical
``patch'' of cylindrical metric functions covering the saddle
point. As discussed in Ref.~\cite{Anninos94a}, the amplitude of the
$\ell=4$ waveform can vary by about a factor two over a wide
range of patch, diffusion, and resolution parameters.  Future refinements
of this code may allow us to make more definitive predictions
of this difficult waveform extraction.

Finally, in Figs.~\ref{figl2mu2.7} and \ref{figl4mu2.7}
we show the $\ell = 2$ and $\ell=4$ waveforms respectively for
the case $\mu = 2.7$, where the holes are well separated by about
$12.6M$ initially. The solid lines are the waveforms extracted at a
distance $r=40M$.
In this case
the fits to perturbation theory are still reasonably good, but are not as
close as the calculations performed for holes that are initially
closer together. The wavelengths of the extracted waveforms are somewhat
too long, and this can be understood as a numerical artifact of our
methods. The calculation must run for a longer period of time
before the onset of quasinormal ringing, so the peak in the radial
metric function becomes more difficult to resolve. This leads to an
error in the longitudinal (spherical) part of the field, that causes
the effective gravitational scattering potential to be somewhat
different from the true potential. (See also Ref.~\cite{Abrahams92a}
for a discussion of this point.) Since this potential is critical in
determining the quasinormal frequencies of the system, the normal
modes are generated at slightly different frequencies.

\subsubsection{Other Radiation Indicators}
\label{subsubsec:indicators}

The Newman-Penrose scalar~\cite{Newman62a}
\begin{equation}
\Psi_4 =R_{\alpha\beta\gamma\delta}l^\alpha \overline m^\beta k^\gamma
\overline m^\delta
\end{equation}
provides another approach that can be used to treat the problem of
radiation extraction.  The vectors ${k}$ and ${l}$ are
orthogonal real
vectors defined
by adding and subtracting a spacelike unit vector with a unit timelike
vector.
The vector ${m}$ and its complex conjugate ${
\overline{m}}$ are orthogonal null vectors tangent to the surface of a
2-sphere representing the wavefront of an outgoing shell of
radiation. The basis set we have chosen to work with is the following
\begin{eqnarray}
k_\mu=&&\frac{1}{\sqrt{2}}\left[-\alpha+\Psi^2\sqrt{A}\beta^\eta,
\ \Psi^2\sqrt{A},\ 0,\ 0\right], \\
l_\mu =&&\frac{1}{\sqrt{2}}\left[-\alpha-\Psi^2\sqrt{A}\beta^\eta,
\ -\Psi^2\sqrt{A},\ 0,\ 0\right], \\
m_\mu =&&\frac{1}{\sqrt{2}}\left[\Psi^2\sqrt{B}\beta^\xi,\ 0,
\ \Psi^2\sqrt{B},\ i\Psi^2\sqrt{D\sin^2\xi}\right],  \\
\overline m_\mu =&&\frac{1}{\sqrt{2}}\left[\Psi^2\sqrt{B}\beta^\xi,\ 0,
\ \Psi^2\sqrt{B},\ -i\Psi^2\sqrt{D\sin^2\xi}\right].
\end{eqnarray}
Far from the source $\Psi_4$
represents an outward propagating wave and is therefore naturally
normal to a 2-sphere of constant ``radius'' $\eta$.
The total radiated energy loss can therefore be estimated
by~\cite{Newman62b}
\begin{equation}
\frac{dE}{dt}=\frac{1}{4\pi} \oint\left[\int_0^t dt' \Psi_4\right]^2
r^2 d\Omega,
\label{nploss}
\end{equation}
where the integration is over a 2-sphere (with an area of $4\pi r^2$
and a surface element $d\Omega$, where
$r\sim\sqrt{\gamma_{\xi\xi}}/M=\Psi^2\sqrt{B}/M$)
lying in a spacelike hypersurface
surrounding the radiating system.

A third method that we have used
to track gravitational radiation is based on
the Bel-Robinson vector \cite{Zakharov73}
\begin{equation}
p^\gamma=E_{\alpha \beta} \epsilon^{\beta \gamma
\delta}B_\delta^\alpha.
\end{equation}
where $E_{\alpha \beta}$ and
$B_{\alpha \beta}$ are the ``electric'' and ``magnetic'' components of the
four dimensional Riemann tensor.
$\varepsilon _\alpha ^{
\gamma \mu} = \varepsilon ^{\gamma \mu}_{\alpha \beta} n^\beta $ is the
3-dimensional Levi-Civita permutation tensor and $n^\alpha$ is the
unit vector normal to the spacelike $t=$ constant hypersurfaces.
Although $p^\gamma$ is constructed in a manner formally similar to the
Poynting vector of electromagnetism, we note that it is not a
physical momentum vector for gravitational waves, since its units
differ by $M^2$.
Nevertheless, $p^\gamma$ has been proven to be effective in
qualitatively tracking gravitational radiation
\cite{Smarr77,Bernstein93a}, and as demonstrated
by DCSE in Ref.~\cite{Smarr79} and again in
section~\ref{subsubsec:energy},
the radiated energies computed using $p^\gamma$
are in good quantitative agreement with those computed using other
radiation indicators.
Since $p^\gamma$ is dimensionally a flux vector quantity, the energy
loss can be approximated by integrating the {\it radial} energy flux
$p^r$ over a closed 2-sphere as described above for $\Psi_4$
\begin{equation}
\frac{dE}{dt}=\frac{1}{4\pi} \oint\left[\int_0^t dt'
\left(\pm\sqrt{\left| p^\gamma r_\gamma \right|/2}\right) \right]^2
r^2 d\Omega.
\label{brloss}
\end{equation}
The choice of sign in the integral of (\ref{brloss}) is taken to be
the sign of $\Psi_4$.  Equation (\ref{brloss}) is motivated by the
asymptotic form $p^{\gamma}r_{\gamma}\rightarrow 2|\Psi_4|^2$
applicable to the case of monochromatic waves in linearized Minkowski
spacetimes~\cite{Smarr77}.  This construction yields results that are
consistent with the integrals of $\psi$ and $\Psi_4$
in the asymptotic far field.

It is informative to compare the waveforms obtained by the three
different methods.
In Fig.~\ref{fig:rad}, we show the $\ell=2$
Zerilli function $\psi$ (solid line), $\Psi_4$
(dashed line) and
$p^r$ (dotted line) at a fixed point $r=70M$
along the equator for the case $\mu=2.2$.  $\Psi_4$ and $p^r$
have been normalized to the same scale of $\psi$ by matching the
amplitudes of their maximum peaks.
The $\ell=2$ fundamental quasinormal mode is clearly present and
dominant in all three signals.

\subsubsection{Energies Radiated}
\label{subsubsec:energy}

The total radiated energy $E$ can be computed from the Zerilli
function using Eq.~(\ref{zloss}). We display these results in
Fig.~\ref{fig:zerillienergy}.  The six clusters of unconnected symbols
represent the six numerical simulations corresponding to the different
$\mu$ parameter values.  Each of the five symbols within a cluster
corresponds uniquely to the total integrated $\ell=2$ energy computed
at the five different wave detectors.  For reference, the early
results of Smarr and Eppley are plotted as large crosses with error
bars suggested by Smarr~\cite{Smarr79}. Within the large errors
quoted, those early results are remarkably consistent with our more
accurate results.

Clearly the results in Fig.~\ref{fig:zerillienergy} show two distinct
regimes, as denoted by the arrows in the upper
part of the figure.  For $\mu < 1.8$ the initial data contains {\em one} black
hole, as discussed in section~\ref{subsubsec:horizons}, and the energy
radiated falls off exponentially.  We have fit the energy
output to an exponential for $\mu \le 1.8$ and find that the results are
approximated well by the formula $E = 3.13 \times 10^{-8} \exp(4.852\mu)$.
For $\mu > 1.8$ there are two holes and the energy radiated is
somewhat independent of the initial separation.  Three lines based on
analytic and semi-analytic calculations that treat the system as two
black holes are also shown, as we discuss in more detail in the next
section.  The curve labelled ``DRPP calculation'' is based on
Ref.~\cite{Davis71}, the result labelled ``Reduced Mass Correction''
takes into account the finite mass of a black hole falling into its
partner, and the dashed line is the complete semi-analytic calculation
discussed in Sec.~\ref{sec:analytic}, accounting for other effects.
Finally we note that the energy radiated is {\em very} small compared
with the upper limits based on the horizon area theorem
(represented by connected circles in Fig.~\ref{fig:zerillienergy}),
as discussed
in the next section.

Table~\ref{table:energy} compares the total radiated energy computed
using three different radiation variables: the Zerilli function
$\psi$, the Newman-Penrose scalar $\Psi_4$ and the Bel-Robinson vector
$p^r$.  The energies in Table~\ref{table:energy} are results from
simulations resolved with a 200$\times$35 grid and are normalized to
the total ADM mass of the spacetime ($2M$).  The range of values
quoted for the numerical calculations for each of the six cases
represent the range of data across the different detector locations,
with the lower (upper) limits corresponding to the outer (inner)
detectors.  As a whole, the results are remarkably consistent.  The
various numerical constructions differ significantly only at the
innermost detectors for the low $\mu$ cases. These deviations are
attributed to several effects. First, the near zones are characterized
by stronger highly distorted behavior than the asymptotic wave zones,
making this region more difficult to resolve accurately. $\Psi_4$ and
$p^r$ are curvature type variables that have an explicit dependence on
first and second order gradients of the metric components. Hence, they
are more susceptible to numerical inaccuracies than the Zerilli
function which depends only on first derivatives of the 3-metric.
Secondly, the curvature quantities are projected onto a coordinate
based tetrad, and the Cadez coordinates are distorted from sphericity
in the inner regions of the grid, obscuring the physical
interpretation of these quantities.

\section{Analytic Estimate}
\label{sec:analytic}

To gain a physical understanding
(and confirmation) of the numerical results, we outline a
procedure to estimate analytically the total
radiated energy. Our approach is based on the well-studied problem
of a test point particle originally at rest at infinity plunging
into a Schwarzschild black
hole~\cite{Zerilli70a,Davis71,Davis72,Detweiler79,Oohara82}.
For the test point particle
problem,~\cite{Zerilli70a}
combined a Newtonian quadrupole moment calculation with the
linearized theory of Landau-Lifshitz~\cite{Landau75}
to find the in-flight radiation
\begin{equation}
E = \frac{1}{105} \left(\frac{m^2}{M}\right),
\end{equation}
for infall from $\infty$ to $r=2M$,
where $m$ is the mass of the test point particle and
$M$ the mass of the black hole with $m\ll M$.
In~\cite{Davis71} the total radiation for the same test point
particle problem is obtained using black hole perturbation
theory (see {\it e.g.,}~\cite{Chandrasekhar83} and references therein).
The result is comparable~\cite{Davis71}
\begin{equation}
E = 0.0104 {m^2 \over M}.
\label{drppenergy}
\end{equation}
Our approach is to adopt the general relativistic
result (\ref{drppenergy}) and modify it to include
correction factors
so that it can describe the
two black hole collision. In the following, we shall discuss
correction factors due to~(A)~$m$ is not much smaller than $M$,
{}~(B)~the infall is not from infinity, and~(C)~the black
hole, unlike a point particle, has a finite size and internal
dynamics.

\subsection{mass scaling}
\label{subsec:mass}

Before we go into the various correction factors, it is useful to
understand why $E$ in~(\ref{drppenergy}) is proportional to
$m^2/M$. For $m \ll M$, the quadrupole moment of the system is $I \sim
m r^2 \quad$ where $r$ is the radial distance between $m$ and $M$. The
gravitational wave luminosity is given by $\overdots{I}$, the third
time derivative of $I$. In the Newtonian approximation, $\dot{r} \sim
\sqrt{{2M / r}}$ and $\ddot {r} \sim {M / r^2}$, we have
\begin{equation}
L \propto \overdots{I}^2
\sim m^2 \bigl(\ddot{r}\dot{r}\bigr)^2
\sim m^2 \biggl( {M^3\over r^5}\biggr) \quad . \label{luminosity}
\end{equation}
The total energy radiated is
\begin{equation}
E =	\int L \, dt
\approx L_{\text {strong\enskip field}}
\times \delta t_{\text {strong\enskip field}} \quad . \label{totale}
\end{equation}
The integral is evaluated at the strong field region, as most energy
is released towards the end when $m$ is falling near the horizon of
$M$. Putting $r=2M$ into~(\ref{luminosity}) for $L$ in the strong
field region, and $\delta t_{\text {strong\enskip field}} \approx M$, we
have
\begin{equation}
E \propto {m^2 \over M},
\label{energy_scaling}
\end{equation}
as in~(\ref{drppenergy}).

When $m$ is not much smaller than $M$,
it is more accurate to use
\begin{equation}
I \sim \mu r^2
\end{equation}
where $\mu \equiv (mM)/(M+m)$ is the reduced mass of the system. Hence
the considerations above (equations~(\ref{luminosity}) to~
(\ref{energy_scaling})) suggest that Eq.~(\ref{drppenergy}) should
be changed to
\begin{equation}
E = 0.0104 {\mu^2 \over M}
\label{4.7}
\end{equation}
as the gravitational wave energy output for the case when $m$ is not
necessarily much smaller than $M$. Notice that for $m=M$, the $\mu^2$
of Eq.~(\ref{4.7}) introduces a quite significant factor of $1/4$.

\subsection{finite infall}
\label{subsec:infall}

The simple quadrupole approximation of~(\ref{luminosity}) also
suggests how the expression should be modified when the infall is not
initially from infinity.  Starting from rest at a finite distance
$r_o$ reduces the velocity $\dot r$, which enters~(\ref{luminosity})
as $\dot r^2$. Denote the energy radiated in such a case as $E_{r_o}$,
\begin{equation}
E_{r_o} = \int_{t_o} L\, dt = \int_{r_o}^{2M} (L/
\dot r )dr \quad .
\end{equation}
Eq.~(\ref{4.7}) is hence modified to be
\begin{equation}
E = F_{r_o} \times 0.0104 {\mu^2 \over M}
\label{4.9}
\end{equation}
\begin{equation}
F_{r_o}
= {E_{r_o} \over E_{\infty}}
= {\int_{r_o}^{2M} \dot{r}(\ddot{r})^2 dr \over \int_{\infty}^{2M}
\dot{r}(\ddot{r})^2 dr } \label{4.10}
\end{equation}
with
\begin{equation}
\dot{r}
= {\biggl( 1- {2M\over r}\biggr) \sqrt{{2M \over r}-{2M \over r_o}} \over
\sqrt{{1-2M\over r_o}} } \quad .
\label{4.11}
\end{equation}
Since $r_o$ can be as small as a few $M$ in the numerical simulation,
we must use the relativistic expression for $\dot{r}$ in Schwarzschild
coordinates in Eq.~(\ref{4.11}). Equations~(\ref{4.10}) and
(\ref{4.11}) represent one way of extending the quadrupole formula to
the highly relativistic regime. There is no unique way to do the
extension, we have just picked a way convenient for our present
purpose. This correction factor $F_{r_o}$ represents two effects:
({\it i})~there is less time to radiate when falling from a finite
distance, and ({\it ii})~the infalling velocity is smaller. The latter
effect is much more important. In Fig.~\ref{fig:Fr0} we plot $F_{r_o}$
vs. $r_o$, the initial Schwarzschild coordinate of $m$, covering the
range of $r_o$ used in the numerical simulation.

\subsection{internal dynamics}
\label{subsec:internal}

All considerations up to this point are the same, independent of
whether the infalling object is a point particle or a black hole. In
the following, we consider correction factors due to this
difference. Although at the end we will extrapolate to two black holes
of equal mass $m=M$, for both generality and convenience of
discussion, we think of the situation as a hole with mass $m$ falling
into a hole with $M \geq m$.

As far as the gravitational wave output is concerned, the most
important difference between a point mass and a black hole is that a
black hole has internal dynamics. There are more channels that the
initial gravitational potential energy in the system can dissipate
into. Such dissipations decrease the kinetic energy and hence the
velocity of the infalling hole. Hence fewer gravitational waves are
generated. There are various mechanisms causing dissipation, which we
shall describe separately. Of course, as these dissipative effects are
more pronounced in the nonlinear regime near the final coalescence, the
separation between the various mechanisms is inevitably of an
approximate nature.

\subsubsection{tidal heating}
\label{subsubsec:tidal}

The first kind of dissipation we consider
originates from the tidal deformation of $m$ as the hole $m$
falls in the static gravitational field generated by $M$.
In the
membrane paradigm~\cite{Thorne86} of black holes, in which the horizon
is treated as a 2-D surface living in a 3-D space, endowed with
physical properties like viscosity, this tidal deformation heats up
the horizon. The heating is described by the horizon
equations~\cite{Thorne86,Suen88,Price86}
\begin{equation} - {d \over
dt} \sigma_{ab} + (g- \theta ) \sigma_{ab} + \bigl( 2\sigma_{ac} +
\gamma_{ac} \theta \bigr) \sigma_b^c = \epsilon_{ab}
\label{4.12a}
\end{equation}
\begin{equation}
- {d \over dt} \theta + g\theta
- {1\over2}\theta^2 =\sigma_{ab}\sigma^{ab} \label{4.12b}
\end{equation}
\begin{equation}
{d\over dt} \gamma_{ab}
= 2\sigma_{ab} + \gamma_{ab}\theta \quad . \label{4.12c}
\end{equation}
Here $\gamma_{ab}$ is the 2-D
metric of the horizon of the infalling hole $m$,
$g=1/(4m)$ is the surface
gravity of the hole $m$, and
$\epsilon_{ab}$
is the normalized electric part of the Weyl tensor ($C_{a\mu b\nu}
l^\mu l^\nu$, with $l^\mu$ the horizon generators).
$\theta$ is the
expansion rate of the horizon generators
\begin{equation}
\theta ={1\over \Delta
A}{d\over dt} \Delta A ,
\end{equation}
and $\sigma_{ab}$ is the shear of the
horizon generators
\begin{equation}
{d\over dt}(\Delta s)^2 =2(\sigma_{ab}
+{1\over 2}\theta\gamma_{ab})\Delta x^a\Delta x^b,
\end{equation}
with $\Delta x^a$
the coordinate separation of the horizon generators.

For a hole with mass $m$ falling
in the external tidal field ${M/r^3}$, $\epsilon_{ab}$ in
an orthonormal
basis can be approximated by~\cite{Thorne86}
\begin{equation}
\epsilon_{ \hat{\theta} \hat{\theta} }
= -\epsilon_{ \hat{\phi} \hat{\phi}}
\sim gV {M \over r^3} \quad ,
\label{4.13}
\end{equation}
where $V$ is the velocity of the fiducial observers on the horizon of
$m$ moving in the external tidal field $(V=\sqrt {2M / r}$ for infall
from $\infty$). Since $\sigma_{ab}$ and $\theta$ in
Eqs.~(\ref{4.12a}-\ref{4.12c}) are driven by
$\epsilon_{\hat{a}\hat{b}}$, which is small in our case, all nonlinear
terms in Eqs.~(\ref{4.12a}-\ref{4.12c}) can be dropped, and
$\gamma_{ab}$ is decoupled. $\theta$ is approximately given by
\footnote{$\theta$ is infinite where caustics exist on the horizon,
although its contribution to the total increase of the horizon area
is finite. We do not consider the effect of caustics in this paper.}
\begin{eqnarray}
\theta
&=& \int \Bigl[ \sigma_{ab} (t^\prime )
\sigma^{ab}(t^\prime )G(t,t^\prime ) \Bigr] dt^\prime \nonumber\\ &=&
\int_{r_o}^{2M}
\Bigl[ \sigma_{ab} \sigma^{ab} G/ \dot{r} \Bigr] dr \label{4.14}
\end{eqnarray}
\begin{eqnarray}
\sigma_{ab}
&=& \int \epsilon_{ab}(t^\prime )G(t,t^\prime )dt \nonumber\\ &=&
\int_{r_o}^{2M} [\epsilon_{ab} G/ \dot{r} ] dr \quad , \label{4.15}
\end{eqnarray}
where $G(t,t^\prime )$ is the teleological Green's
function~\cite{Thorne86}
\begin{equation}
G(t,t^\prime ) = \left\{
\begin{array}{l}
\exp [g(t-t^\prime )] \; \mbox{for}\quad t < t^\prime \\ 0 \quad\qquad\qquad
\mbox{for}\quad t > t^\prime \label{4.16}
\end{array}\right.
\end{equation}
In Eqs.~(\ref{4.14}) and (\ref{4.15}), $\dot{r}$ is given by
Eq.~(\ref{4.11}); the integrands are regarded as functions of $r$,
with $t=t(r)$ obtained by integrating Eq.~(\ref{4.11}). In
Eqs.~(\ref{4.14}) and (\ref{4.15}) the integrations are cut off when
the hole falls through the horizon of $M$. In principle the
integration should be carried over all times, but the residue is
unimportant for the present purpose. In Fig.~\ref{fig:shear}, we plot
$\sigma_{\hat{\theta}\hat{\theta}}$ (solid line) and $\theta$ (dotted
line) as functions of the Schwarzschild coordinate $r$, for the case
of $m=M$. The horizontal and vertical scales are in terms of
$M=m=1$.

A fraction of the initial gravitational potential energy $f_h
=\Delta m/m$ is dissipated into the heating of the horizon of $m$ by
this effect:
\begin{eqnarray}
f_h = {\Delta m \over m} &=& {1\over 2} {\Delta A\over A} =
{1\over 2} \int \theta \, dt \nonumber\\ &=& {1\over 2}
\int_{r_o}^{2m+2M} [\theta / \dot{r} ] dr \label{4.17}
\end{eqnarray}
Notice that the integration in Eq.~(\ref{4.17}) is terminated at the
point when the two holes are engulfed by a common horizon. It is
irrelevant whether the object that has fallen in is a point mass or a
black hole after that point. We approximated that point to be when the
holes are separated by $2m+2M$ in Schwarzschild coordinates, {\it i.e.,}
when the two holes are nearly touching. For $m=M$, the heating on the
horizon of $M$ is the same as that on $m$, hence the total fraction of
energy going into heating of the horizons as the holes are falling in
each others tidal field is given by $2f_h$. In
Fig.~\ref{fig:Fr0}, the reduction factor for the energy
available for wave generation,
\begin{equation} F_h =1-2f_h
\label{4.18}
\end{equation}
is plotted against $r_o$, the initial separation of the holes, for the
case of $m=M$. We see that $F_h$ is decreasing with increasing initial
separation, as a larger initial separation leads to a larger velocity
and in turn larger $\epsilon_{\hat a \hat b}$ in Eq.~(\ref{4.13}). As
$r_o\rightarrow \infty$, $F_h$ decreases to 0.86. For the range of
$r_o$ covered in the numerical simulation, this effect reduces the
gravitational wave output by about 10\%.

\subsubsection{absorption of gravitational waves}
\label{subsubsec:absorption}

The second kind of dissipation arises from the fact that, unlike a
point mass, a black hole has finite size. As it sweeps through the
spacetime, it can reabsorb the gravitational wave already generated in
the spacetime.

The gravitational wave that a black hole can absorb depends on the
frequency $\omega$ of the wave and the $l-pole$ of the wave:
\begin{equation} {\text {wave\enskip absorbed}} = \sum_l \int (l- {\text
{pole\enskip wave\enskip incident\enskip on}}\enskip m ) \times T_l
(\omega ) d\omega \, .
\label{4.19}
\end{equation}
Here $T_l$ is the transmission coefficient for incident $l$ pole waves
as calculated in black hole perturbation theory, see
{\it e.g.,}~\cite{LIGO3}.
We take the $l$-pole wave incident on the infalling black hole
$m$ as the $l$-pole wave in the spacetime times the cross-section of
the hole $m$, as seen from the hole $M$, {\it i.e.,}
\begin{equation}
{\text cross\enskip section} \lesssim {\pi (2m)^2 \over 4\pi (2m+2M)^2} = {1
\over 4} {\mu^2 \over M^2}
\label{4.20}
\end{equation}
The energy in the $l$-pole wave in the spacetime is given {\it e.g.,}
in Ref.~\cite{Thorne94a},
and is reproduced in Fig.~\ref{fig:lpoleE} (dashed line) for
$l=2$ as a function of $\omega$. The vertical axis (for the dashed
line) is $dE/d\omega$, all in scale of $M=1$. The transmission
coefficient $T_{l=2}$ times 0.1 is also given as a function of
$\omega$ (dotted line). Notice that the quasi-normal frequency is
$0.3737 m^{- 1}$. $T_{l=2}$ is about 0.5 at this point, dropping to
zero rapidly for a smaller $\omega$. As the peak of $dE/d\omega$ is at
a smaller $\omega$, the product of $T\, dE/d\omega$ is small, given by
the solid line in Fig.~\ref{fig:lpoleE} for the case of $m=M$. The
area under the curve is found to be $\sim 0.0012 m$. Comparing to the
total gravitational wave energy in the $l=2$ mode~\cite{Davis71},
$E_{l=2} = 0.0092 m$, Eqs.~(\ref{4.19}) and (\ref{4.20}) lead to about
1\% reabsorption of the $l=2$ wave energy.

Reabsorption of higher $l$ modes can be estimated similarly. However,
since the peak of $dE_l/d\omega$ is always at an $\omega$ less than
the corresponding quasi-normal frequency $\omega_l$, whereas the
transmission coefficients rise to larger than 0.5 only for $\omega >
\omega_l$, the reabsorption is always a small fraction of the
corresponding component. As the $l=2$ mode is the dominant component
making 90\% of the total radiation, we see that this reabsorption
effect can only lower the total energy output at a 1\%
level \footnote{As the effect of reabsorption is small, various refinements
of equations~(\ref{4.19}) and~(\ref{4.20}) are not meaningful, {\it e.g.,}
the cross-section~(\ref{4.20}) is in fact a function of the separation
between the two holes and the $\ell=2$ mode of the two black hole
system is different from the $\ell=2$ mode of the infalling
hole of mass $m$.}
\begin{equation}
F_{\text {abs}} \approx 99\% .
\end{equation}

We point out that there are other mechanisms causing the
heating of the black hole horizons during the coalescence of black
holes. Indeed, some of them are much larger than the effects
considered above. For example, in the late stages of the coalescence,
the infalling hole $m$ is moving relativistically and is beaming
gravitational waves in the forward direction. Such beaming causes large
horizon heating~\cite{Suen88,Price86}. The energy dissipated is an
order of magnitude larger than the total gravitational wave output to
infinity. However, such an effect is already implicitly included in
the result of~\cite{Davis71}. Hence no modification factor is
needed to account for the process. The reduction in gravitational wave
output due to the interference from the different parts of a body for
a finite size object falling into a black hole has been estimated in
Ref.~\cite{Haugan82}.
For the case of a two black hole collision, the effect is
negligible.

\subsection{comparison with numerical results}
\label{subsec:comparison}

In Fig.~\ref{fig:zerillienergy} we plot the final result of the
analytic estimate for total gravitational wave energy output
\begin{equation}
\label{totalenergy}
E=F_{r_o}F_hF_{\text{abs}}
\times 0.0104 {\mu^2 \over M}
\label{4.21}
\end{equation}
versus the initial separation between the two holes, for the case of
$m=M$. It is represented by the dashed line. For comparison to the
final result of Eq.~(\ref{4.21}) we have also plotted two intermediate
results as straight, solid lines.  A simple application of the standard
perturbative ``DRPP calculation'' (Eq.~\ref{drppenergy}) overestimates
the energy by roughly an order of magnitude.  Replacing $m$ by the
reduced
mass $\mu$ as in Eq.~(\ref{4.7}) gives the ``Reduced Mass
Calculation'', accounting for the finite mass of the
``perturbing'' black hole.  The three correction factors $F_{r_o}$,
$F_h$, and $F_{\text{abs}}$ together  act to reduce the energy
output further, leading to the curved, dashed line in
Fig.~\ref{fig:zerillienergy}.  In view of the various approximations
one has to make to obtain Eq.~(\ref{4.21}), the agreement between the
analytic and the numerical results is remarkable, as the analytic
results were obtained without prior knowledge of the numerical
results, and vice versa. For $L/M$ less than about 9 the analytic
formula overestimates the actual energy output computed
numerically. This is to be expected since for small enough separations
the holes are initially engulfed by a common effective potential
(in the sense of the potential in the black hole perturbation
theory), or even a common event horizon. The
approximation for colliding black holes in these
cases is inappropriate.

The connected circles show
the maximum possible radiation output obtained by comparing the
initial black hole masses estimated by the areas of the horizons (or a
single horizon if the holes are close enough) to the total mass of the
spacetime. For large separations this number approaches 29\% as
expected from the work of Hawking~\cite{Hawking73}.

What physical understanding is gained from this semi-analytic exercise?
The central message is that, as far as gravitational radiation is
concerned, a black hole falling into another black hole is
not much different from that of a point particle falling into
a black hole. As the leading order approximation, the energy
output can be described quite well by Eq.~\ref{4.9}, which is
the test point particle result obtained in~\cite{Davis71} modified
by insight from the quadrupole formula. We found that the biggest
effect due to black hole internal dynamics can be understood as the
deformation of the horizons when the holes are falling in the
tidal fields of one another. We see that the energy of about
$10^{-3}M$ (for black holes of mass $M$) is dissipated by the
viscosity of the horizon. Instead of radiating out to infinity, this
portion of the initial potential energy is dissipated into the holes,
increasing the horizon area. The smallness of this number again
testifies that the horizon of a black hole is ``stiff'', making a
black hole rather like a point particle.
At present we can only say the contribution from horizon heating
is consistent with our numerical results to the level of accuracy.
A future direct confirmation of this effect, especially for the
case of non-equal mass black holes, will be particularly interesting.

There are other effects that we considered, for example
the reabsorption of radiation (also interference
calculated in~\cite{Haugan82})
due to the finite size of the black hole. We found these effects
to be negligible beyond the present level of
accuracy, adding further weight to the understanding of a black hole
behaving to a large extent as a point particle. There are
also other effects
that we have thought about but do not know yet how to calculate
analytically, for example the increase of the horizon area of the
black hole due to caustics. However, the agreement between the
numerical results and our present semi-analytic
approximations suggests that
these other effects most likely do not affect the total
energy radiated significantly.

\section{Conclusions}
\label{sec:conclusions}

We have performed numerical and analytic calculations predicting the
gravitational waveforms generated and total gravitational wave energy
emitted when two equal mass black holes collide head-on. Waveforms for
all cases studied show similar behavior: the normal modes of the final
black hole are excited and account for most of the emitted
signal. Both the $\ell=2$ and $\ell=4$ waveforms are fit nicely
by a superposition of the fundamental and first overtone of the
black hole quasinormal modes. Although the fit to perturbative
calculations of the $\ell = 4$ waveform is quite good, the amplitude
and precursor of this waveform are more sensitive to the various
computational parameters than the more dominant
$\ell=2$ waveform.

The total energy radiated is on the order of $0.002M$ (where
$M$ is half the ADM mass of the spacetime),
far below the estimate given
by a simple application of the area theorem.
The analytic study,
appropriate for holes that are initially separate, confirms and
elucidates the numerical results.
We find the total energy radiated can be approximated quite well
using the point particle result modified slightly to account
for mass scaling, finite initial separation and internal
dynamics of the black holes.
Taken together, the
analytic and numerical results indicate that even for holes that are
initially infinitely separated, the total energy output will be the
same order of magnitude.  For throats that are close together, Price
and Pullin~\cite{Price94a} have treated the evolution via
gauge-invariant perturbation methods, regarding
the system as a single, perturbed hole.  They find remarkable
agreement with our work for the $\ell = 2$ waveforms and total energy
radiated, independently
confirming our results and providing a new method for
evolving distorted black hole data sets.  A detailed comparison of the
numerical results presented here to semi-analytic results based on
regimes where the throats are initially near or far from each other
will be published elsewhere.

The work presented in this paper is only a first step on the long path
to computing the fully general evolution of
three-dimensional, spiralling, coalescing
black holes. We expect the more general case to be significantly more
complex to compute. In axisymmetry our current calculations can be
extended to include: boosted black hole collisions in which the
holes are given initial finite velocities,
unequal mass black holes
which can radiate not only gravitational waves but net linear
momentum, and
spinning colliding black holes where one can expect
more energy to be radiated, particularly if the holes have opposite
spin vectors. We intend to pursue such extensions to our present
axisymmetric code and to develop and apply new, more general,
three dimensional codes~\cite{Anninos94c} to these
systems as well.

Finally, we note that we have prepared a video showing results for
several of the simulations reported here. Interested readers may
contact NCSA media services at the internet address
media@ncsa.uiuc.edu for information on how to obtain a copy of the
video entitled ``The Collision of Two Black Holes.''  At this address
one can also obtain a copy of a videotape of the original movies based
on the work of Smarr and Eppley.  The NCSA group has also set up a
World Wide Web server accessible at the URL
http://jean-luc.ncsa.uiuc.edu, and there one can find images and
movies of the simulations presented in this paper that cannot be
published in traditional form.

\acknowledgements
We would like to thank David Bernstein for a number of helpful
discussions, Joe Libson and Paul Walker for integrating light rays
through the spacetimes, Richard Price and Jorge Pullin for informing
us of their results prior to publication, Mark Bajuk for his work on
visualizations of our numerical simulations that aided greatly in
their interpretation, and Joan Mass\'o for help with preparing some of
the graphs for this paper. This work was supported by NCSA, NSF
Grant~91-16682, and NSERC Grant No. OGP-121857, and calculations were
performed at NCSA and the Pittsburgh Supercomputing Center.



\newpage

\begin{figure}
\caption{
A single quadrant of the Cadez grid is displayed for the case $\mu=2.2$.
The throats are centered on the
axis at $z=\pm \coth\mu$.
Lines of constant $\eta$ concentrically surround the throats locally,
and become spherical far from the holes.}
\label{fig:cadezgrid}
\end{figure}

\begin{figure}
\caption{
The metric function $\hat\gamma_{\eta \eta}=A$
for the case $\mu = 1.2$ is shown
at coordinate time $t = 25M$, where $M$ is the mass parameter defined
to be half the ADM mass. This configuration was surrounded by a global
apparent horizon initially.}
\label{grr1.2}
\end{figure}

\begin{figure}
\caption{The lapse function $\alpha$ is shown for the case
$\mu = 1.2$ at the
time $t = 25M$. It collapses to zero uniformly around the entire region
surrounding both throats, indicating that the system is behaving as a
single black hole.}
\label{lapse1.2b}
\end{figure}

\begin{figure}
\caption{The lapse function $\alpha$ is shown for the case $\mu = 3.25$
at time $t = 22.5M$.  At this point the two black holes are still evolving
independently of each other.}
\label{lapse3.25}
\end{figure}

\begin{figure}
\caption{The mass of the apparent horizon is shown for the
cases $\mu = 1.2$ (solid line)
and $\mu=2.2$ (dashed line).
In the $\mu=1.2$ case the initial horizon surrounds both
holes and most of the mass-energy is contained within the
horizon. After a slight increase early in the evolution (within the
first few $M$) the mass of the horizon remains essentially constant
until $t \sim 30M$ when it begins to grow due to numerical effects
discussed in the text.  In the $\mu=2.2$ case the initial
horizons around each throat are distinct. At $t \sim 17M$ a new
apparent horizon appears, surrounding both holes and accounting for
essentially all the mass-energy in the system.}
\label{fig:horizon}
\end{figure}

\begin{figure}
\caption{The area of the constant $\eta$ surface that just encircles
the saddle point is plotted as a function of $\mu$. The solid line
is $16\pi$, the area corresponding to the total ADM mass of the
spacetime (we work with units such that the total mass is normalized
to unity).}
\label{fig:hoop}
\end{figure}

\begin{figure}
\caption{
We plot various timings
of the evolution for the six initial data
sets. $t_{newt}$ is the Newtonian free fall
time required for two particles to collide
from rest at this separation, $t_{hor}$ is the time at which the
apparent horizons merged, and $t_{ring}$ is the time at which the
first (negative) peak is seen in the $\ell = 2$ Zerilli function
recorded at $r = 40M$.}
\label{fig:time}
\end{figure}

\begin{figure}
\caption{
The $\ell = 2$ waveforms for the case $\mu=1.2$. The solid
line is the numerically generated waveform extracted at $r=40M$.
The dashed line is
a fit of the two lowest $\ell = 2$ quasinormal modes, over
the domain $70<t/M<160$, to the extracted waveform.}
\label{figl2mu1.2}
\end{figure}

\begin{figure}
\caption{
The $\ell = 2$ waveforms for the case $\mu=2.2$.
The solid line is the waveform extracted
at $r=40M$ and the
long dashed line is the waveform at $r=60M$. The dotted line is
the quasinormal mode fit.}
\label{figl2mu2.2}
\end{figure}


\begin{figure}
\caption{
The $\ell = 4$ waveforms for the case $\mu=2.2$.
The solid line is the waveform extracted at $r=40M$
and the short dashed line is the quasinormal mode fit.}
\label{figl4mu2.2}
\end{figure}

\begin{figure}
\caption{
The $\ell = 2$ waveforms for the case $\mu=2.7$.
The solid line is
the waveform extracted at $r=40M$
and the short dashed line is the quasinormal mode fit.}
\label{figl2mu2.7}
\end{figure}

\begin{figure}
\caption{
The $\ell = 4$ waveforms for the case $\mu=2.7$.
The solid line is the
waveform extracted at $r=40M$ and the short dashed line is the
quasinormal mode fit.}
\label{figl4mu2.7}
\end{figure}

\begin{figure}
\caption{
The time evolutions for three different radiation indicators
are shown for the case $\mu=2.2$.
The solid line is the Zerilli function ($\ell=2$),
the dashed line is the Newman-Penrose scalar $\Psi_4$, and
the dotted line is the radial component of
the Bel-Robinson vector.
The data tracks the behavior in time of a single point
on the grid located at
$r=70M$ along the equator.
The $\ell = 2$ fundamental quasinormal mode
is clearly present and dominant in all three signals.}
\label{fig:rad}
\end{figure}

\begin{figure}
\caption{
The total gravitational wave energy output is shown for the six
parameter studies.  The connected circles are the upper limit based on
the area theorem, the clustered symbols show numerical results at
various detector locations, and the crosses show early results by
Smarr and Eppley with their approximate error bars.  The solid line
labelled ``DRPP Calculation'' is the result of a naive application of
the point particle result, the solid
line labelled ``Reduced Mass Correction'', takes into account the
finite mass of the infalling black hole, and the dashed line is the
semi-analytic estimate, including several effects discussed in the
text.  Finally, the dot-dashed line is an empirical fit to the data
for low values of the separation parameter $\mu$, showing the
exponential falloff of the energy in the one black hole regime.}
\label{fig:zerillienergy}
\end{figure}

\begin{figure}
\caption{
We plot the energy
factors $F_{r_o}$ and $F_h$ vs. $r_o$. The horizontal axis is the
initial Schwarzschild coordinate $r_o$ of the infalling hole
in units of $M$.
$F_{r_o}$ is the gravitational wave energy
output for infalling from $r_o$ divided by that from
$\infty$. $F_{r_o}$ is smaller for small $r_o$ mainly because the
velocity of the infall is smaller.
$F_h$ is the percentage of the available energy compared to the total
gravitational energy available after taking into account the
dissipation due to tidal distortions of the black hole horizons.
As $r_o \rightarrow \infty$, $F_h$ tends to 86\%.
The curves are calculated
for $m=M$.}
\label{fig:Fr0}
\end{figure}

\begin{figure}
\caption{
Development of horizon shear and expansion. The horizontal axis is the
Schwarzschild coordinate $r$ of the infalling object in units of
$M$. The solid line represents the magnitude of the horizon shear on
the horizon of the infalling hole $m$. The dotted line is the
expansion. Both the shear and the expansion are in units of $m^{-1}$
and are plotted for the case $m=M$.}
\label{fig:shear}
\end{figure}

\begin{figure}
\caption{
The reabsorption of the $l=2$ wave. The horizontal axis is the angular
frequency $\omega$ of the $l=2$ wave in units of $M^{-1}$. The dashed
line is $dE/d\omega$, the dotted line is the
transmission coefficient of the $l=2$ wave
multiplied by 0.1 and the solid line is
the product $T(dE/d\omega )$.
For $M=m$ the area under the solid curve, which is found to be
$0.0012m$, is roughly the total reabsorption of the $l=2$ wave by the
infalling hole.}
\label{fig:lpoleE}
\end{figure}

\begin{table}
\begin{tabular}{|c|c|c|c|c|} \hline
$\mu$	& $M$ & $L/M$ & Apparent horizon & Event horizon \\ \hline
1.2 & 1.85 & 4.46 & global & global \\ \hline
1.8 & 0.81 & 6.76 & separate & critical \\ \hline
2.2 & 0.50 & 8.92 & separate & separate \\ \hline
2.7 & 0.29 & 12.7 & separate & separate \\ \hline
3.0 & 0.21 & 15.8 & separate & separate \\ \hline
3.25 & 0.16 & 19.1 & separate & separate \\ \hline
\end{tabular}
\caption{The physical parameters for the six initial data sets.
$M$ is the mass parameter equal to half the ADM mass of
the spacetime, $L/M$ is the proper
distance between the throats, and we note whether or not
a single apparent
or event horizon surrounds both holes.
\label{table:horizon}}
\end{table}

\begin{table}
\begin{tabular}{|c|c|c|c|} \hline
$\mu$ & $E_{\psi}$ & $E_{\Psi_4}$ & $E_{p^\gamma}$ \\ \hline 1.2 &1.34 - 0.99
$\times 10^{-5}$
&9.81 - 1.32 $\times 10^{-5}$
&7.58 - 1.29 $\times 10^{-5}$ \\ \hline
1.8 &1.95 - 1.64 $\times 10^{-4}$
&3.46 - 1.72 $\times 10^{-4}$
&3.77 - 1.73 $\times 10^{-4}$ \\ \hline
2.2 &6.10 - 5.27 $\times 10^{-4}$
&7.76 - 5.21 $\times 10^{-4}$
&8.91 - 5.26 $\times 10^{-4}$ \\ \hline
2.7 &7.50 - 6.86 $\times 10^{-4}$
&7.16 - 5.00 $\times 10^{-4}$
&7.52 - 5.07 $\times 10^{-4}$ \\ \hline
3.0 &8.85 - 7.13 $\times 10^{-4}$
&9.22 - 4.86 $\times 10^{-4}$
&10.0 - 4.93 $\times 10^{-4}$ \\ \hline
3.25&1.37 - 0.85 $\times 10^{-3}$
&4.32 - 1.11 $\times 10^{-3}$
&3.39 - 1.11 $\times 10^{-3}$ \\ \hline
\end{tabular}
\caption{The total radiated energy for the six initial data sets
normalized to the ADM mass ($2M$) of the spacetime.
We compare results using several different methods of calculation on a
grid resolved with 200$\times$35 zones.  The energies are extracted at
five different
radii, and the range given here is from the innermost to
the outermost radius.
See text for details.
\label{table:energy}}
\end{table}

\end{document}